# Quantitative Single-Shot Supercontinuum-Enhanced Terahertz Spectroscopy (SETS)

**Authors**

Charan R. Nallapareddy,[1] Thomas C. Underwood[1,2]*

**Affiliations**

[1]Department of Aerospace Engineering and Engineering Mechanics, The University of Texas at Austin, TX 78712.
[2]Texas Materials Institute, The University of Texas at Austin, TX 78712.
* Corresponding Author. Email address: thomas.underwood@utexas.edu

**Abstract**

Single-shot terahertz (THz) spectroscopy probes sub-picosecond, non-repetitive events by combining the advantages of laser absorption techniques with the phase-sensitive detection of interferometric methods. However, its usage as a quantitative tool is hindered by the chirp penalty of the spectral encoding scheme, where a narrow probe bandwidth distorts the THz signal, limiting the bandwidth and spectral resolution of THz measurements. In this work, we introduce Supercontinuum-Enhanced Terahertz Spectroscopy (SETS), a method that leverages a broadband supercontinuum probe to overcome these challenges. With SETS, we show an increase in usable THz bandwidth from 1.5 THz to 2.3 THz, reducing signal distortion by 50% and offering a scalable pathway to extend bandwidth further. Numerical models of spectral encoding highlight the flexibility of SETS, achieving high spectral resolution (< 4 GHz with a 250 ps supercontinuum pulse) while maintaining a usable bandwidth > 1 THz for amplitude and phase spectra. Experiments on argon plasma and water vapor, complemented by theoretical validation, generalization, and extrapolation, show the capability of SETS to measure electron density, collision frequency, and absorption spectra simultaneously with tailored measurement accuracy and resolution (as low as $5\times10^{15}$ m$^{-3}$ for electron density). By addressing a critical limitation in single-shot diagnostics, SETS enables high-resolution, non-intrusive, quantitative measurements for complex reactive flows and dynamic refractive media.

**Keywords**

Single Shot, Terahertz Time Domain Spectroscopy, Chirp Penalty, Dispersion Penalty, Supercontinuum, Plasma, Absorption Spectroscopy, Interferometry, Gas Sensing

## 1. Introduction

Revealing the dynamics of ultrafast, non-repetitive events—such as phase transitions,[1] shapes of relativistic electron beams,[2,3] combustion processes,[4,5] and plasma physics[6]— requires optical processes that are sensitive and adaptable across a range of characteristic timescales. Single-shot terahertz time domain spectroscopy (THz-TDS), a pump-probe technique,[7] enables sub-picosecond measurements while capturing dynamics at the laser repetition rate. It also combines the strengths of traditional spectroscopy, such as laser absorption measurements, with the capabilities of interferometric methods like microwave interferometry[8] but with higher spatial resolution (due to the shorter wavelength). This versatility stems from its time domain measurement approach, which can be transformed into the frequency domain using Fourier analysis, providing both amplitude and phase spectra. Despite these advantages, quantitative measurements with the technique are challenging—particularly capturing an amplitude and phase spectrum with high resolution (i.e., spectral) and a wide bandwidth.[9,10] This limitation arises from a fundamental constraint, the "chirp/dispersion penalty"[2], that is inherent to single-shot methods that use a temporally-chirped probe pulses to encode the signal of interest (i.e., the THz electric field). As a probe pulse with fixed bandwidth must encode the entire THz time domain signal in single-shot schemes, spectral encoding methods face a trade-off that is analogous to a narrow range of frequences that attempt to carry too much information. This results in distortion due to an insufficient capacity in a probe pulse to represent a THz signal fully.

Extensive research has been conducted to overcome the chirp penalty, including approaches that are distinct from spectral encoding, such as echelon-based[11–14] and angular encoding techniques[15]. However, these methods present their own challenges, such as the inability to capture transverse spatial information of the medium in the test section—a capability that is inherent to the spectral encoding technique.[16,17] There have been other techniques involving chirped-pulse cross-correlation characterizations that require complex designs and sensitive alignment,[11,18] however they lead to new trade-offs.[2] While advanced techniques like phase diversity[2] offer improvements over the traditional spectral encoding scheme and can capture spatial information, THz signals with pulse widths approaching the probe's transform-limited pulse width cannot be measured without significant distortion. This is where supercontinuum probe pulses that feature an adjustable bandwidth can become useful.

In this work, we employ a supercontinuum to expand the bandwidth of a probe pulse and overcome the chirp penalty in single-shot THz-TDS that uses spectral encoding. We compare the spectrally modulated chirped pulse signal with a benchmark delay line signal to quantify the impact of the chirp penalty in the time domain for the first time (Fig. 1B-1C) as a distortion factor, β. We highlight how the penalty restricts the spectral resolution and bandwidth in the frequency domain (Fig. 1D-1E) due to the formation of artifacts and null frequencies, which hinder quantitative spectroscopy. To address these challenges, we introduce *Supercontinuum-Enhanced Terahertz Spectroscopy (SETS)* – a new class of ultrafast diagnostics for measuring the refractive index of dynamic media (Fig. 2) that is designed to reduce the distortion factor. This method overcomes constraints that relate the usable THz bandwidth, the spectral resolution of a THz waveform, and the chirped pulse width in single-shot measurements. The use of a supercontinuum probe pulse offers a mechanism to eliminate the formation of null frequencies while expanding the usable THz bandwidth beyond what is permitted with a chirped pulse width that features a limited bandwidth.

Measurements were performed to demonstrate the phase accuracy of THz-TDS measurements using both delay-line and single-shot spectral encoding methods (e.g., chirped and SETS), comparing the time (Fig. 3) and frequency domain (Fig. 4-6) content of a chirped pulse (1.9 ps, 22 nm) and a supercontinuum pulse (2 ps, 45 nm) (Fig. 2, Materials and Methods). Our measurements indicate that a supercontinuum pulse in SETS reduces β by 200%, extends the usable THz bandwidth from 1.5 to 2.2 THz, and maintains a spectral resolution of 0.3 THz (Fig. 3). We generalize and extrapolate these results using a spectral encoding model (Materials and Methods, SI Secs. 1, 5, 7) that simulates probe beam modulation by THz waveforms in an electro-optic crystal and signal demodulation at the spectrometer, revealing the potential for spectral resolution < 4 GHz with a 250 ps supercontinuum pulse (Fig. 4). The advantages of SETS for quantitative spectroscopy are demonstrated in gaseous and plasma systems, both examples of refractive media that can vary in space and time. Using a gaseous cell and a radio-frequency plasma source, we validate the linearity and accuracy of refractive index and gaseous measurements inferred from absorption (e.g., gaseous absorbance) and phase shifts (e.g., plasma density) (Materials and Methods, SI Sec. 5-6). Increasing the probe bandwidth by 10-fold reduces the uncertainty in $H_2O$ concentration measurements by 8-fold, while increasing the probe pulse width enhances the resolution for measuring electron densities within plasmas from $10^{17}$ m$^{-3}$ (2 ps pulse) to $5 \times 10^{15}$ m$^{-3}$ (100 ps pulse) (Fig. 5). These results highlight SETS as a sensitive, high-resolution tool for refractive index measurements, with performance driven by factors like phase uncertainty, linearity, and probe pulse properties.

## 2. Results

### 2.1 The problem: chirp/dispersion penalty

Spectral encoding is a technique that is employed commonly to extract THz-TDS in a single shot of a laser pulse.[9–11] This method modulates a THz time domain signal onto the frequency components of a chirped probe beam at an electro-optic crystal (Fig. 1A). The THz modulated probe beam is then sent into a spectrometer to extract the THz signal (SI Sec. 1) by mapping the frequency components of a probe beam to the time domain using a time-wavelength calibration (Materials and Methods, SI Sec. 4). This method, however, introduces a chirp penalty because of mismatches that exist between the THz time domain information and the finite bandwidth of a probe pulse (i.e., measurement window vs. probe bandwidth). A finite bandwidth can only accommodate a limited amount of information while maintaining a phase-accurate representation of the original modulating signal (i.e., the THz electric field) after demodulation. However, a narrow probe bandwidth is often chirped to longer pulse durations and forced to carry the time domain information from extended THz pulses, which, while increasing the measurement window of the diagnostic, leads to signal distortion and a lack of phase accuracy.

Quantitative applications of THz-TDS require a high spectral resolution, typically up to ~ 10 – 100 GHz, to resolve line shapes and measure phase shifts across a broad range of frequencies. Achieving this level of resolution necessitates single-shot approaches that are capable of measuring THz signals over an extended time period, often spanning ~10-100 ps, to ensure sufficient sampling for a given resolution and sensitivity. For example, extending the measurement time (i.e., chirped pulse width, $T_c$) in THz-TDS resolves lower THz frequencies ($1/T_c$), which enhances the sensitivity for detecting refractive media that exhibit a Drude response (Fig. 5C).[19] Decreasing the spacing between measurements in the time domain improves the resolution in resolving absorption features (Fig. 4C-E). This is due to the reciprocal relationship between spectral resolution in the Fourier domain ($\delta\omega$, the spacing between two adjacent data points in the frequency domain) and the measurement duration in the time domain, which is governed by the number of time domain steps, the signal-to-noise ratio (SNR) of the diagnostic, and the spacing between successive measurements. However, as chirped pulse widths increase to improve the spectral resolution, the distortion of the THz signal increases, shrinking the usable bandwidth ($\Delta\omega$) in the frequency domain (Fig. 1B). This distortion creates a trade-off between $\delta\omega$ and $\Delta\omega$ in single-shot THz-TDS, restricting the quantitative accuracy of the diagnostic (Fig. 1B).

The extent of signal distortion can be quantified using the distortion factor, $\beta$ (SI Sec. 1-2, Fig. 1C), where lower $\beta$ values indicate that a signal extracted in a single shot more closely resembles an undistorted THz signal (pulse width, $T_{THz}$, Fig. 1A). $\beta$ is proportional to the probe bandwidth (or its transform-limited pulse width, $T_0$) and $T_c$, where a narrower probe bandwidth (larger $T_0$) or longer $T_c$ increases signal distortion. The effect of β on a THz waveform can be evaluated analytically in the limit of chirp rate (α) $\gg \frac{1}{T_c^2}$ and $T_c^2 \gg T_{THz}^2$ (SI Sec. 1) or numerically using fast Fourier transforms (FFTs) (Materials and Methods, SI Sec. 2) and wavelength-time conversion. In this work, we adopted the numerical approach because it can be applied to any THz waveform, including those that are measured experimentally, not just well-defined mathematical forms like the bipolar signal (Eq. 3). Additionally, while the analytical expression (Eq. S2.4, SI Sec. 2) captures the stretching effect of the distorted signal, it fails to capture the multi-cyclic nature (as shown in Fig. 1C) of the distortion.[20] Capturing these features in the time domain using a numerical approach (Fig. 1C) is essential because they introduce artifacts in the frequency domain (i.e., null frequencies) which limit Δω in both amplitude (Fig. 1D) and phase (Fig. 1E) spectra. The higher the β, the higher the number of null frequencies in the frequency domain. The presence of these features limits the quantitative accuracy of the diagnostic. While shortening $T_c$ can reduce β, it also limits δω which can constrain the range of frequencies over which phase shifts can be measured or the ability to resolve absorption features in an amplitude spectrum. For example, a β=8 corresponds to a narrower Δω (~ 0.5 THz) compared to β=0.8, and thus fails to capture a spectral feature of interest present at 1 THz (Fig. 1D) (i.e., a bipolar signal devoid of a 1 THz sine wave, Fig. 1C). To improve and optimize performance, a quantitative single-shot diagnostic needs to extend the measurement window to enhance δω while also minimizing β. Since $T_{THz}$ is fixed for a given generation scheme, reducing $T_0$ of a probe pulse (i.e., increasing its bandwidth) becomes one method that is used in SETS to reduce β and expand Δω without compromising δω.

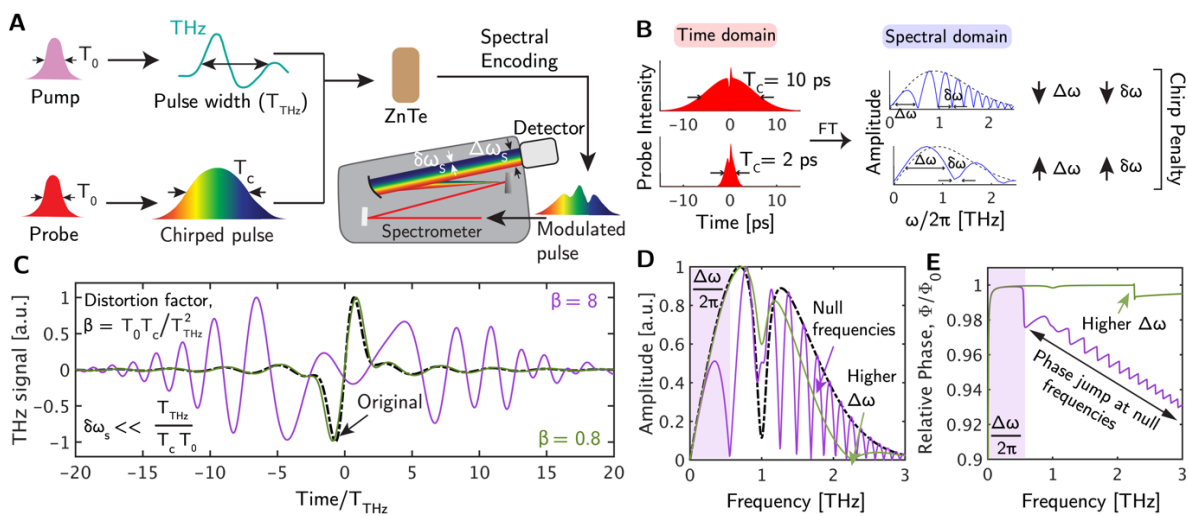

**Figure 1. Spectral encoding technique and the impact of the chirp penalty.**
**(A)** The limited bandwidth of a chirped probe pulse allows the extraction of THz time domain signals in the spectral domain using a spectrometer. This measurement introduces a chirp penalty. That is **(B)** for a given probe bandwidth, increasing the chirped pulse width ($T_c$) improves spectral resolution (δω), due to the Fourier transform relationship between the time and frequency domains, but reduces the usable bandwidth (Δω) due to introduction of null frequencies. **(C)** This chirp penalty manifests as signal distortion (characterized by β) in the time domain and imposes a trade-off between Δω and δω in both **(D)** amplitude and **(E)** phase spectra in the frequency domain. This tradeoff restricts the applicability of chirped THz-TDS for quantitative measurements.

## 2.2 A solution: Supercontinuum-Enhanced Terahertz Spectroscopy (SETS)

Supercontinuum pulses[21,22] (Materials and Methods) provide a broadband probe that overcomes a fundamental limitation of single-shot measurements by expanding the probe's bandwidth independently of its chirped pulse width. In this study, three methods for measuring THz radiation in a TDS configuration were implemented: a delay line method, a chirped pulse single-shot method with a narrow and limited probe bandwidth, and a chirped supercontinuum single-shot method with adjustable probe bandwidth (i.e., SETS) (Fig. 2A, Materials and Methods). The diagnostic setup employed a ytterbium femtosecond laser to pump an optical parametric amplifier (OPA), producing a separate 1532 nm, 80 fs pump beam and a 780 nm, 350 fs probe beam, each with ~1 µJ pulse energy and repetition rates of up to 2 MHz. THz radiation was generated using the pump beam via optical rectification in a 500 µm-thick PNPA organic crystal, achieving >4% efficiency,[23] and detected using a 2 mm-thick <110> ZnTe crystal that modulated the probe beam through the electro-optic effect, responding proportionally to the local THz electric field (Materials and Methods). In delay line detection, the polarization-modulated probe pulse was converted to amplitude modulation using a Wollaston prism and a quarter-wave plate in a balanced photodetection setup, with data acquired at 133 fs/step using a motorized delay stage, lock-in amplifier, and digital oscilloscope. Single-shot detection used chirped probe pulses that were polarization-modulated using electro-optic modulation, converted to amplitude modulation with cross-polarizers (extinction ratio >1000:1), and analyzed using a spectrometer equipped with an EMCCD camera (Materials and Methods). This modular design supported multiple detection methods while ensuring high SNR. For SETS, a supercontinuum was generated by focusing the probe beam onto a 6 mm thick sapphire crystal. By adjusting the focal point of the probe within the crystal, the probe's bandwidth was controlled independently of the chirp pulse width.

The delay line method, while offering high accuracy and temporal resolution, is limited by its slow mechanical scanning process, making it unsuitable for capturing transient reactive flow phenomena.[24] This method introduces a trade-off between temporal resolution (set by the delay line stepping and $T_0$ of the probe) and the total measurement time window (set by the scanning range). In this work, the delay line signal is used for benchmarking purposes and validating the single-shot results in terms of temporal, amplitude and phase accuracy. The chirped pulse method with limited bandwidth, though faster with measurements times of a single-shot, succumbs to the chirp penalty from spectral encoding, leading to time domain signal distortion. In contrast, the chirped supercontinuum single-shot method overcomes this limitation by generating a broadband supercontinuum pulse compared to a narrow probe bandwidth for the same temporal pulse width. Experimentally, we demonstrated a 2-fold increment in bandwidth with supercontinuum for a fixed temporal pulse width of ~ 2 ps (Fig. 2B). This approach reduces $\beta$ by increasing the carrier/probe bandwidth, preserves high $\delta\omega$, and extends $\Delta\omega$, offering a more efficient and accurate method for capturing THz radiation with enhanced fidelity.

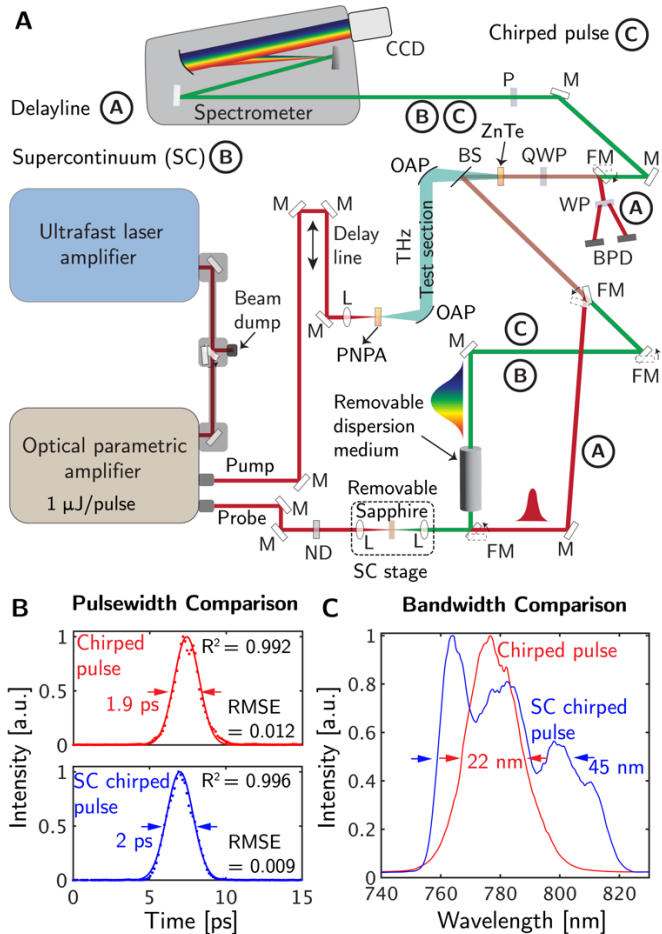

**Figure 2. Single-shot THz detection schemes and a solution to the chirp penalty using a supercontinuum.**
**(A)** A schematic illustrating three THz detection schemes: a conventional delay line technique, which raster scans the THz time domain signal using multiple probe pulses to extract different points sequentially; a single-shot approach via spectral encoding using a chirped pulse with a limited bandwidth; and a single-shot spectral encoding method using a supercontinuum, which expands the bandwidth of a probe pulse using a sapphire crystal. Key components in the setup of the SETS diagnostic include, OAP: off-axis parabolic mirror, M: mirror, L: lens, FM: flip mount, BS: beam splitter, P: polarizer, ND: neutral density filter, QWP: quarter-wave plate, WP: Wollaston prism, and BPD: balanced photodetectors. **(B)** The supercontinuum pulse is chirped to 2 ps using a single pass of a 5 cm SF11 rod, in contrast to the original probe beam, which requires a 10 cm SF11 rod to achieve a similar chirp duration. **(C)** This difference arises from the ~ 2-fold higher bandwidth of the supercontinuum pulse. *(single-column figure)*

## 2.3 A supercontinuum improves THz pulse extraction

In this section, experiments and theory were developed to understand the root cause of distortion and validate the spectral encoding model with experimental data from SETS. Experiments were performed to demonstrate how increasing the bandwidth of a probe beam using a supercontinuum pulse, or equivalently reducing $T_0$, reduces the distortion in extracted THz signals (Figs. 3A-B). For instance, experimental measurements showed a regular chirped pulse with limited bandwidth (22 nm) resulted in a β of 1.22, which improved to 0.64 when switching to a supercontinuum pulse with a bandwidth that is 2 times larger (45 nm). This demonstrates that a supercontinuum pulse more accurately reconstructs the original delay line signal (obtained through point-by-point sweeping of the delay line, Materials and Methods) compared to a chirped pulse. Additionally, simulations were performed using the spectral encoding algorithm (SI Sec. 1, Materials and Methods) to replicate the electro-optic modulation and demodulation of chirped ($T_0$ = 40 fs) and supercontinuum ($T_0$ = 20 fs) signals, assuming a modulation constant, k, of 0.01 (|k|≪1, Materials and Methods). These simulations used the delay line THz electric field, measured experimentally, as the undistorted THz waveform. The simulations were used to validate the experimental results and extrapolate the results further in quantitative analysis of the amplitude and phase spectra.

The 2-fold reduction in distortion that was observed in the experimental time domain data (Figs. 3A-B) enhanced spectral characteristics, increasing $\Delta\omega$ from 1.5 THz to > 2.2 THz while maintaining a fixed $\delta\omega$ of 0.3 THz. This improvement is evident in both the amplitude (Fig. 3C) and phase spectra (Fig. 3D), showcasing how a supercontinuum pulse enhances the overall spectral fidelity. In amplitude spectra, $\Delta\omega$ is constrained by the location of the first null frequency, marking the effective limit of usable bandwidth. In the phase spectra, represented as a ratio relative to the benchmark delay line phase ($\phi_0$), deviations from unity indicate the influence of null frequencies and their disruptive effect on the spectral data as discussed in Figs. 1D and 1E. Therefore, when the spectral information of the target THz electric field extends beyond the first null frequency, the time domain signal becomes distorted. This is because null frequencies act as a virtual "low-pass filter," removing or perturbing critical high-frequency components (i.e., those beyond the first null frequency). These null frequencies get closer to each other progressively with $T_c$ (Fig. 1D) and thereby remove a significant portion of the spectrum beyond the first null frequency, leading to issues with quantitative absorption spectroscopy (Fig. 4) and phase linearity (Fig. 5). The higher the β, the narrower the frequency range that this low-pass filter allows, leading to increased distortion. Thus, minimizing β is essential to preserve the time domain signal and its spectral accuracy.

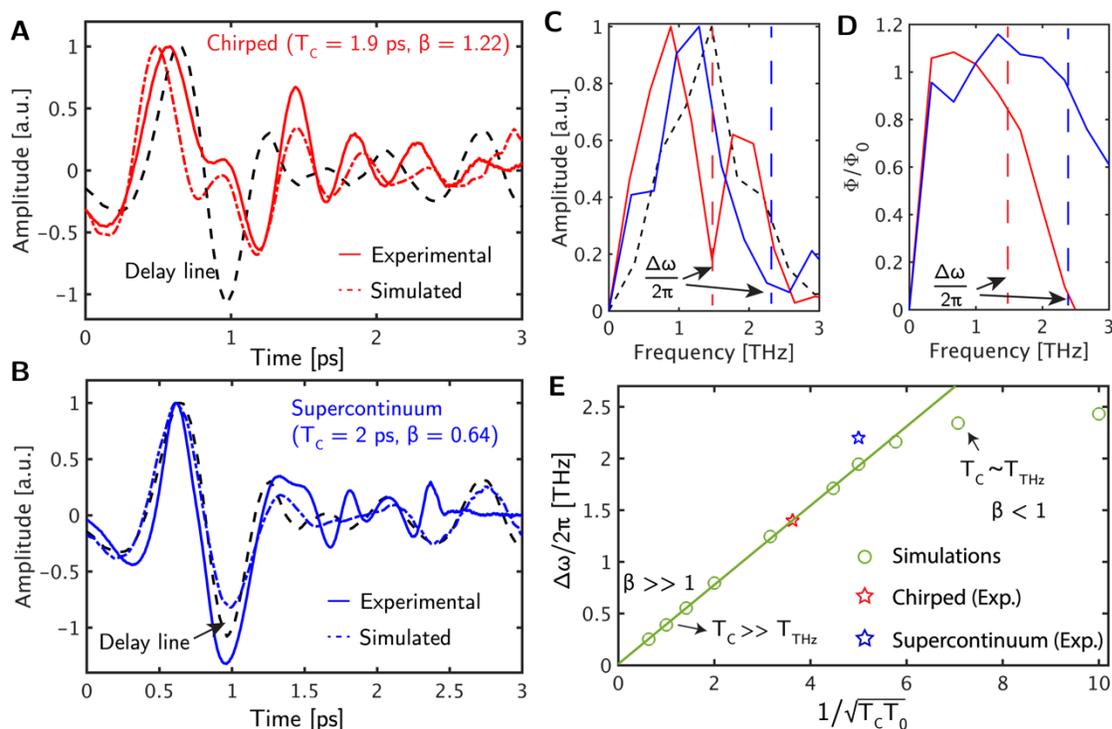

**Figure 3. A supercontinuum probe pulse reduces THz distortion and enhances the accuracy of spectral features.**
**(A)** Experimental and simulated results showing that increasing the bandwidth of the probe beam (or reducing $T_0$) decreases the distortion factor in the time domain (quantified using the relation for β shown in Fig. 1C and developed in SI Sec. 2), reducing it from β = 1.22 with the chirped pulse to **(B)** β = 0.64 with the supercontinuum pulse. This improvement is reflected as increased $\Delta\omega$ for a fixed $\delta\omega$ in both **(C)** the amplitude spectrum and **(D)** the phase spectrum (represented as a ratio relative to the delay line phase, $\phi_0$) for experimental measurements. **(E)**

The bandwidth can be tuned further by adjusting $T_0$ relative to $T_c$. However, as $T_c$ approaches $T_{THz}$, deviations from the ideal behavior (in the limit of $T_c \gg T_{THz}$) occurs.

Tuning $T_0$ relative to $T_c$ allows control over $\beta$ and enables adjustments to $\Delta\omega$ and $\delta\omega$ in single-shot approaches (Fig. 3E), manipulating the characteristics of the virtual "low-pass filter". While increasing $T_c$ improves $\delta\omega$, it narrows $\Delta\omega$. For example, for a $T_c$ of ~ 10 ps, $\delta\omega$ changes from ~ 0.3 THz (corresponding to a $T_c$ of ~ 2 ps) to ~ 0.06 THz. However, this increase in $T_c$ reduces $\Delta\omega$ from ~ 1.5 THz to < 0.75 THz (Fig. 3E). To counteract this, a proportional 5-fold reduction in $T_0$ maintains a consistent low $\beta$, preserving the original $\Delta\omega$, because $\Delta\omega \propto \frac{1}{\sqrt{T_c T_0}}$. Nevertheless, reducing $\beta$ allows $\Delta\omega$ to expand but only up to the intrinsic THz bandwidth or the phase-matching limit of the electro-optic crystal, beyond which it plateaus to this limit (Fig. 3E).

## *2.4 Impact of supercontinuum probing on the amplitude spectrum*

In this section, experiments were performed to quantify the gaseous absorption features using single-shot THz-TDS. Theory was developed to extend experimental results to higher $T_c$ values using simulated THz pulses, both in the presence and absence of water vapor, to dictate conditions that are necessary for a desired accuracy in gas sensing applications. By applying a Beer's law fit[25] to their Fourier counterparts (Materials and Methods, SI Sec. 7), we investigate the impact of freely adjustable $T_0$, enabled through the use of a supercontinuum, on the accuracy of quantitative absorption spectroscopy. Experiments demonstrated how optimizing both the experimental setup and measurement conditions is crucial for quantitative THz spectroscopy, particularly in extracting physical properties like the line-averaged number density (ñ). This involved analyzing the amplitude spectra of the THz signal that was modulated by the presence of water vapor (i.e., I) and the background THz signal without water vapor absorption (i.e., $I_0$). The spectra were obtained by applying an FFT to the corresponding time-domain THz signals (Materials & Methods). Calculating the logarithm of the ratio of these spectra results in the absorption spectrum, which was then fit to determine ñ. Measurements were performed using a purge box maintained at 1 atm to isolate the THz beam path of 70 cm from the surroundings and control the water vapor content (Fig. 4A) or relative humidity (RH) within a range of <1% to ~14%. While supercontinuum showed an increase in $\Delta\omega$ (Fig. 3C), $\delta\omega$ of 0.3 THz remained inadequate for quantitative absorption spectroscopy due to the use of a 2 ps probe pulse. Applying Beer's law to absorbance data (Materials and Methods, SI Sec. 7) obtained from a 2 ps chirped and supercontinuum-extracted THz signals resulted in fits that extracted ñ with > 80% uncertainty due to a large $\delta\omega$ of 0.3 THz (Fig. 4B).

We simulated THz absorption by water vapor for various measurement windows (i.e., $T_c$ in the case of single-shot methods) to understand the impact of $T_c$ on the absorption spectrum. In these simulations, the original delay line signal passes through a section filled with humid air at an RH of ~14%, allowing us to determine the theoretical undistorted THz signal modulated by water vapor. This signal is then transformed into the single-shot time domain signal using spectral encoding (detailed in SI Sec. 7). These simulations show that, by expanding $T_c$, $\delta\omega$ is reduced and more THz time-domain information is captured. For instance, when $T_c$ is increased to ~10 ps, $\delta\omega$ improves from ~0.3 THz (at $T_c$ ~ 2 ps) to ~0.06 THz. Therefore, increasing $T_c$ is a pathway to improve absorption spectrum data for a better fit and extraction of ñ (Fig. 4C). In this work, the accuracy of the fit is quantified as the ratio of estimated ñ from the fit for a given $T_c$ to the actual ñ content in the purge box as measured by a thermal conductivity humidity sensor. Nevertheless, increasing $T_c$ also introduces a chirp penalty, which can be mitigated by reducing $T_0$ prior to applying this approach for quantitative absorption spectroscopy.

While increasing $T_c$ captures more THz information and improves the accuracy of spectral absorption fits, the reliability of the fit is also sensitive to $T_0$ (Figs. 4C-D) due to the chirp penalty from spectral encoding. For example, at $T_c = 2$ ps, experimental data reveals that supercontinuum pulses with a 45 nm bandwidth result in approximately a 22% improvement in fitting the J-branches of $H_2O$ between 0.4 and 1.1 THz, compared to chirped pulses with a more limited bandwidth of 22 nm. This discrepancy arises because higher $T_0$ values in chirped pulses introduce more null frequencies, as seen in Fig. 4C. These nulls can overlap with critical peaks in an absorption spectrum, amplifying uncertainty in both the spectral fit and the extracted ñ. In contrast, reducing $T_0$ minimizes null frequencies, improving the reliability of the fit (Fig. 4D). This highlights the importance of optimizing $T_0$ alongside extending $T_c$ for accurate measurements. We demonstrate this by fitting two absorption spectra corresponding to a fixed $T_c$ of 250 ps (i.e., 4 GHz of spectral resolution) but with different probe bandwidths ($T_0$) (Fig. 4E). The spectrum extracted with higher $T_0$ shows null frequencies as expected and a few of them overlap with the peaks of interest, increasing the uncertainty in the extracted ñ (or relative humidity in this case, as the medium was humid air). On the other hand, the spectrum that was simulated with lower $T_0$ exhibits no null frequencies, enabling the use of a larger $T_c$ to reduce δω while mitigating the associated increase in β. This results in improved spectral fidelity and reduced measurement uncertainty, which was previously not possible or shown with single-shot THz-TDS systems.

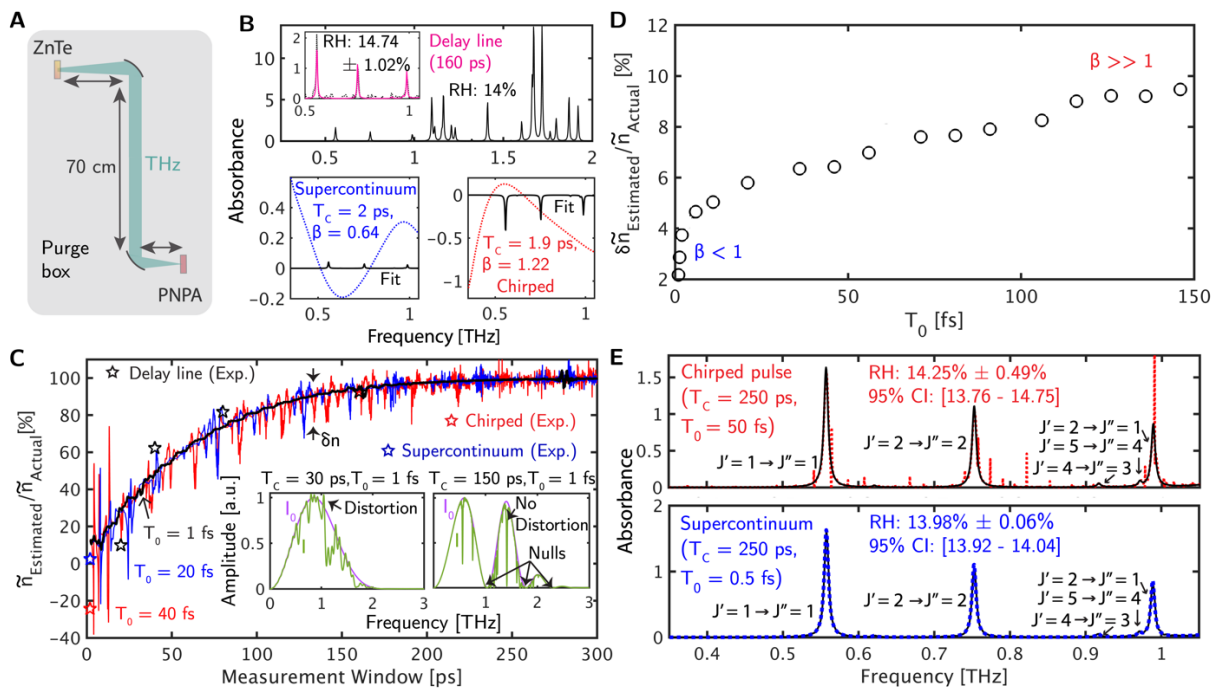

**Figure 4. A supercontinuum enables quantitative absorption spectroscopy in single-shot measurements.**
**(A)** A schematic of the experimental setup and purge box used for gas absorption measurements. **(B)** Experimental data plots with fits for delay line, chirped pulse, and supercontinuum signals, showing unreliable fits for supercontinuum and chirped pulses due to a short measurement window of 2 ps. **(C)** Expanding the measurement window (i.e., the chirped pulse width, $T_c$) improves the accuracy of the fits for extracting the line-averaged number density in single-shot measurements (i.e., concentration) of gaseous molecules (ñ=$\frac{1}{L}\int_0^L\int$ ndL, where n is the number density and L is the path length). However, **(D)** the uncertainty in estimating ñ increases as $T_0$ increases (i.e., as the probe bandwidth decreases). The uncertainty (δñ) was quantified by calculating the standard deviation of the ñ vs measurement window curve for a given $T_0$, after subtracting its moving average to remove the baseline variation. **(E)** Accurate fits of H$_2$O rotational lines are achieved in the limit of long $T_c$ and short $T_0$ for single-shot schemes. A bootstrap statistical analysis[26] gave the confidence intervals of these fits.

## 2.5 Impact of supercontinuum probing on the phase spectrum

In this section, experiments are performed to demonstrate the measurement of refractive properties of media using phase shifts in single-shot THz-TDS. A combination of theory and simulations are applied on an ideal bipolar signal (Eq. 3) to establish how a supercontinuum is advantageous compared to chirped pulses with limited bandwidth at extracting a phase spectrum (Fig. 5). We then measure phase spectra using SETS on a plasma as a refractive medium of interest[9,10,27–30] to test the quantitative capabilities of SETS and validate the claims (Fig. 6, SI Sec. 6).

Two critical aspects of phase spectra are phase linearity, $\delta\omega$, and $\Delta\omega$. Phase linearity refers to the frequency range over which the phase changes linearly with frequency, and this linear range determines the usable bandwidth ($\Delta\omega$). Increasing the chirped pulse width $T_c$ enhances $\delta\omega$ but reduces $\Delta\omega$, as the increase in $\beta$ disrupts phase linearity (or close to unity in the case of $\phi/\phi_0$) due to the presence of null frequencies. We show this by simulating a spectrally encoded single-shot time domain signal from an ideal bipolar signal (Eq. 3, $T_{THz}$ = 0.25 ps) and transforming it into frequency domain using FFT to find the phase spectrum (Fig. 5A). The null frequencies introduce abrupt changes in $\phi/\phi_0$ (Fig. 5A), mirroring their impact on the absorption spectrum. A supercontinuum helps mitigate this by reducing $\beta$, maintaining high spectral resolution and preserving phase linearity across a broader frequency range. For example, for a $T_c$ of 2 ps, a 2-fold increase in bandwidth due to the supercontinuum, reduces $\beta$ by half and increase $\Delta\omega$ from ~ 1.5 THz to ~ 2 THz. As discussed earlier for absorption spectrum, increasing $T_c$ (to say 10 ps from 2 ps, Fig. 5A) limits $\Delta\omega$ due to the chirp penalty, which a supercontinuum overcomes.

On the other hand, understanding the need for higher $\delta\omega$ originates from the need to infer refractive properties using a phase shift ($\Delta\phi$) spectrum. [9,10,23–26] For example, when a THz waveform passes through a plasma, an example of a medium that has a refractive index varying as a function of frequency, it incurs a phase shift ($\Delta\phi$). This is because different frequency components of the waveform travel at different speeds, resulting in signal dispersion. A higher $\delta\omega$ in $\Delta\phi$ spectra, or a longer $T_c$, is advantageous for the quantitative characterization of a plasma for a few reasons: 1) to access low-frequency data such as collision frequency ($\nu_{en}$) (Fig. 5B), and 2) to improve resolution for measuring low plasma electron density ($\tilde{n}_e$) or plasma frequency ($\omega_p$) (Fig. 5C). These properties can be inferred from a measured phase shift using the Drude model, which relates the plasma properties like $\omega_p$ and $\nu_{en}$ with $\Delta\phi$ (Materials and Methods, SI Sec. 5) through a plasma's complex refractive index (N). Therefore, single-shot THz spectroscopy can serve as a valuable diagnostic tool for measuring $\omega_p$ (or $n_e$) and $\nu_{en}$. Since $\Delta\phi$ drops with frequency, a higher $\delta\omega$ allows for measuring larger $\Delta\phi$ values that exceed its experimental measurement uncertainty (SI Sec. 8). Thus, longer $T_c$ enables the detection of lower $\tilde{n}_e$ (Fig. 5C). For example, increasing the probe pulse width from 2 ps to 100 ps enhances the $\tilde{n}_e$ resolution from $10^{17}$ m$^{-3}$ to $5\times10^{15}$ m$^{-3}$ respectively (Fig. 5C). Further increasing $T_c$ can extend this range even more, but $T_0$ must be reduced proportionally to keep $\beta$ low and minimize signal distortion. A limitation of this diagnostic is that when $\nu_{en}$ is large, the $\tilde{n}_e$ resolution quickly plateaus and no longer improves with increasing $T_c$ (Fig. 5C). This means that the diagnostic's ability to measure progressively lower $\tilde{n}_e$ diminishes when $\omega_p/\nu_{en}\ll1$, as the maximum achievable $\Delta\phi$ is constrained by $\omega_p/\nu_{en}$ (Fig. 5B).

The THz electric field undergoes both amplitude attenuation and phase shift when interacting with a dispersive plasma. This behavior arises from reflection losses at the plasma-air interface and the dispersive nature of plasma. To extract plasma parameters such as $\omega_p$ and $\nu_{en}$, it is necessary to analyze both amplitude and phase changes for a given frequency, similar to microwave interferometry.[8] However, unlike interferometric techniques, THz-TDS enables simultaneous broadband measurements in the time domain, allowing direct comparison of undispersed and plasma-dispersed signals. This provides two methods to extract properties of a plasma: 1) *Amplitude and phase analysis:* By applying the Drude model, both $\omega_p$ (or $n_e$) and $\nu_{en}$ can be extracted simultaneously, forming a well-defined two-variable system. This can also be achieved by fitting THz time domain data obtained experimentally to simulated plasma-dispersed THz signal (Materials and Methods, SI Sec. 5-6). 2) *Phase-shift analysis:* For conditions where $\omega \gg \nu_{en}, \omega_p$, the simplified Drude model can extract $\tilde{n}_e$ independently from $\Delta\phi$ spectra (Materials and Methods, SI Sec. 5).

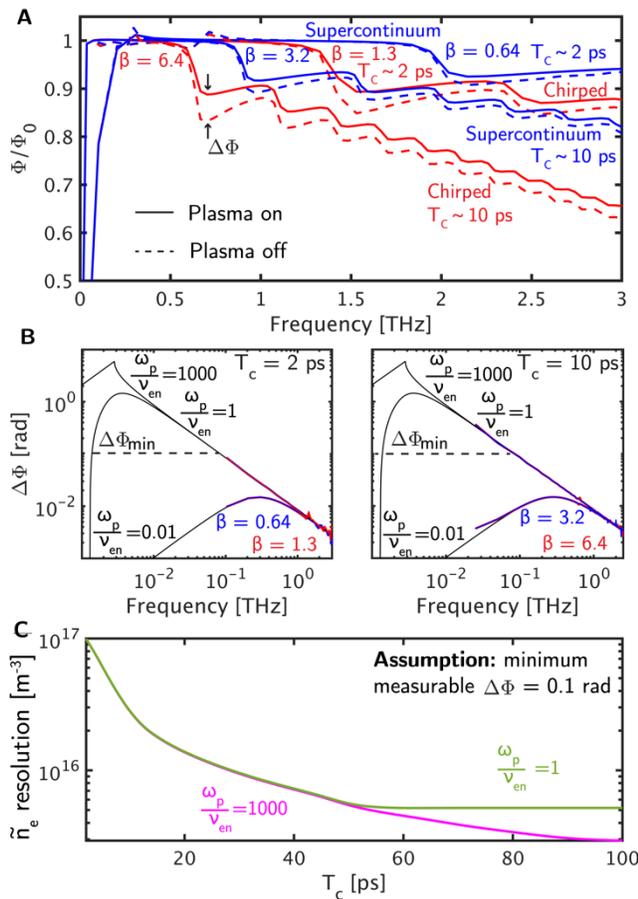

**Figure 5. A supercontinuum enhances phase linearity and enables the measurement of lower free electron number densities.**
Simulated results demonstrating that **(A)** the supercontinuum increases the usable bandwidth ($\Delta\omega$) compared to a chirped pulse of the same duration ($T_c$), ensuring phase linearity without null frequencies (i.e., non-linearity) over a broader range of frequencies. **(B)** The supercontinuum allows $T_c$ extension while maintaining a lower $\beta$ compared to a chirped pulse, facilitating the detection of phase shifts ($\Delta\phi$) at lower frequencies. **(C)** Consequently, lower electron number densities ($\tilde{n}_e$) can be measured, determining the minimum measurable (or $\tilde{n}_e$ resolution). As the ratio of plasma frequency ($\omega_p$) to collision frequency ($\nu_{en}$) decreases, $\tilde{n}_e$ resolution plateaus with increasing $T_c$, as $\Delta\phi$ from the Drude model reaches a global maximum at higher frequencies, preventing further growth of $\Delta\phi$ at lower frequencies. A longer $T_c$ also provides access to lower frequency data, enabling the measurement of properties such as $\nu_{en}$. The $\tilde{n}_e$ used in simulating the results shown in (A) and (B) is $1\times10^{17}$ m$^{-3}$. Pathlength, L, is 10 cm. The minimum measurable $\Delta\phi$ of 0.1 rad was from the phase measurement uncertainty of the experimental setup (SI Sec. 8). *(single-column figure)*

Experiments were conducted using a 27.17 MHz inductively coupled plasma at 220 mTorr, with RF power varied from 150 W to 500 W (Fig. 6A) to demonstrate the use of SETS as a diagnostic to infer the refractive properties of a plasma. Single-shot time domain measurements were conducted using chirped (22 nm) and supercontinuum (45 nm) pulses. The experimental results were compared to simulated THz pulse shapes (Fig. 6B) by applying the spectral encoding algorithm (SI Sec. 1) to the delay line signal (Fig. 6B) and compare with the experimental results, corresponding to a reduction in β from 1.22 (chirped) to 0.64 (supercontinuum). To extract $\tilde{n}_e$, the time domain data was transformed into the frequency domain using a Fourier Transform (Materials and Methods) to analyze $\Delta\phi$, the phase shift induced by the plasma. This approach is valid when ω (~1 THz) >> $\nu_{en}$ (~ 6.5 GHz at 150 W) and $\omega_p$ (~ 4.5 GHz at 150 W). The values for $\nu_{en}$ and $\omega_p$ were determined by transforming the undispersed original experimental THz waveform into the frequency domain, applying the plasma transfer function, and iteratively minimizing the least square difference between the simulated and experimental signals (Materials and Methods, SI Sec. 5-6). Finally, the extracted $\Delta\phi$ from the experiments was fitted to the simplified Drude model, which relates $\Delta\phi$ directly to $\tilde{n}_e$ (Materials and Methods, SI Sec. 5), to determine $\tilde{n}_e$ (Fig. 6C). The single-shot results were validated against baseline delay line data, with measurements repeated at different powers (250 W, 330 W, 400 W, and 500 W) for all three measurement types. The $\tilde{n}_e$ values obtained were $2.4 \times 10^{19}$, $2.9 \times 10^{19}$, $3.3 \times 10^{19}$, and $3.8 \times 10^{19}$ m$^{-3}$ respectively. This confirmed a linear increase in $\tilde{n}_e$ with RF power (Fig. 6D, SI Sec. 9).

Chirped pulses showed earlier deviations in $\Delta\phi$ compared to supercontinuum pulses due to a higher β (1.22 compared to 0.64, respectively). This indicates that $\Delta\omega$ for chirped pulses is shorter (1.4 THz) than that of supercontinuum pulses ($\Delta\omega > 2$ THz). The broader $\Delta\omega$ of supercontinuum pulses expands the bandwidth of the 'virtual low-pass filter,' allowing more frequencies of the time domain THz waveform to pass through. This shifts the null frequencies to higher frequencies, preserving phase linearity and enabling $\tilde{n}_e$ extraction over a wider range of frequencies (Fig. 6C). The use of a supercontinuum also enables larger $T_c$ values to be used while maintaining the same THz bandwidth, features that enable lower frequencies to be resolved and more sensitive detection of plasmas (i.e., phase shifts increase for the same refractive properties at lower frequencies in plasmas). This observation aligns with the experimental results in Fig. 2, which demonstrate measurement consistency as long as β remains unchanged.

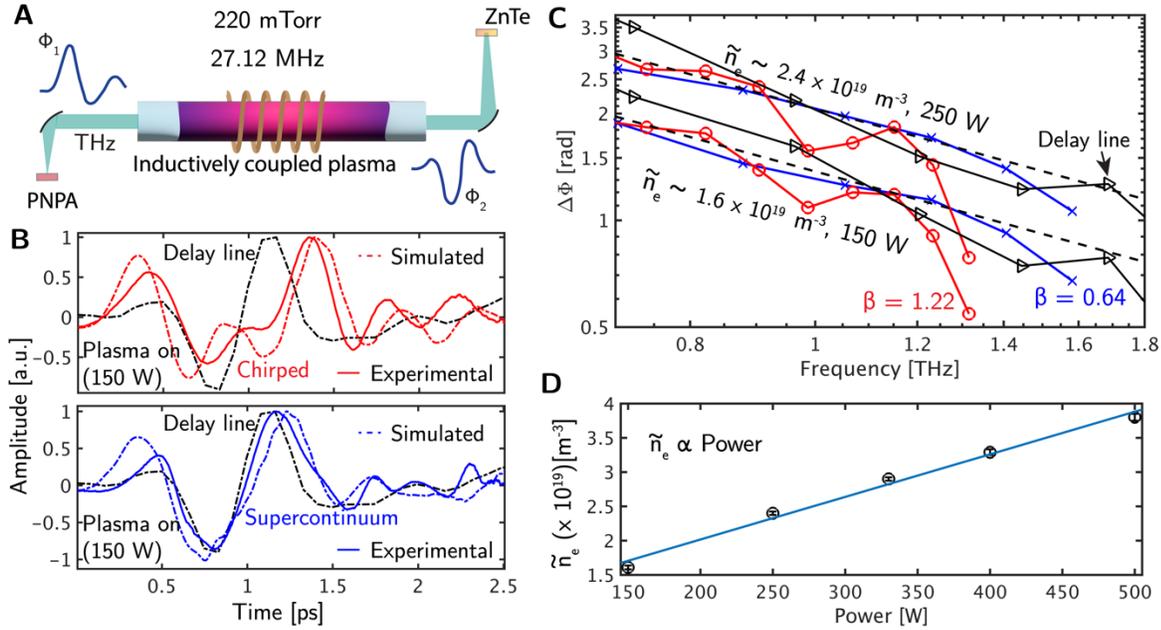

**Figure 6. SETS as a single-shot quantitative plasma diagnostic.**
**(A)** Schematic of an inductively coupled plasma flow reactor that was used to measure Δϕ under varying RF powers at a fixed pressure of 220 mTorr and frequency of 27.12 MHz. **(B)** Experimental and simulated time domain THz signals transmitted through a plasma with an input power of 150 W RF demonstrate that the supercontinuum approach matches the delay line method closely, outperforming the chirped pulse in accuracy. **(C)** This improvement expands Δω for a more sensitive quantitative extraction of $\tilde{n}_e$. **(D)** An increase in the applied RF power led to a linear rise in $\tilde{n}_e$, demonstrating a direct proportionality between $\tilde{n}_e$ and RF power. Details about the linearity at low RF powers (or low ionization fraction) can be found in SI Sec. 9. The raw time domain signals used to extract $\tilde{n}_e$ as a function of RF power are shown in SI Sec. 10.

## 3. Discussion

This study establishes SETS as an effective solution to the limitations that are inherent in single-shot THz diagnostics that use spectral encoding. By addressing the "chirp penalty" using broadband supercontinuum pulses, SETS overcomes long-standing trade-offs between the spectral resolution and bandwidth of THz. Our findings demonstrate that SETS improves the fidelity of single-shot THz diagnostics and make it useful for quantitative measurements, extending the usable spectral bandwidth from 1.5 THz to over 2.2 THz while maintaining low δω. Using spectral encoding algorithm, these results were further extrapolated and generalized to a broader range of $T_c$ (> 250 ps) and $T_0$ (< 1 fs).

SETS enables quantitative spectroscopy by minimizing null frequencies and reducing β, as demonstrated through experimental measurements and theoretical simulations. This advancement supports accurate and reliable measurements of the refractive properties of dynamic media. Additionally, SETS achieves rapid sub-picosecond measurement timescales, resolving dynamics at the laser repetition rate, making it ideally suited for investigating non-repetitive and ultrafast events. We studied plasma as a dynamic medium and measured critical properties of its refractive index, such as $\omega_p$ (or $n_e$) and $\nu_{en}$. The supercontinuum also enabled high-resolution absorption spectroscopy, which was previously not possible, to extract species number densities with desirable accuracy (i.e., depending on $T_c$). The ability to optimize $T_0$ and $T_c$ balances spectral resolution and phase linearity, enhancing measurement reliability over a wider frequency range. Although challenges remain, such as the plateau effect in $\tilde{n}_e$ resolution at high $\nu_{en}$, this approach shows potential for advancing THz diagnostics and enabling deeper understanding of dynamic and refractive media, with applications extending to material science and fusion research.

## 4. Materials and Methods

### *4.1 Diagnostic Design*

*Terahertz Time Domain Spectroscopy (THz-TDS)*: The THz-TDS system consisted of a THz generation module using optical rectification, modulation of a probe beam using the electro-optic effect, and a flexible modular design for THz detection. The design supported a transition between three experimental setups: a traditional delay-line-based THz-TDS, single-shot acquisition with a chirped pulse of finite bandwidth, and single-shot acquisition with a chirped supercontinuum pulse. Each setup was used to compare the THz transmission through gaseous media in a test section.

*Amplified Femtosecond Laser System:* A Spectra-Physics Spirit 1030-70 ytterbium femtosecond laser (pulse energy > 70 µJ) was used to pump a Spectra-Physics NOPA VISIR OPA. The amplified signal (1532 nm, 80 fs) served as the pump beam to generate THz, while the idler (780 nm, 350 fs) was used as the probe beam to detect THz in both a delay line and single-shot configuration. Both the pump and probe beams were polarized vertically, with pulse energies of ~ 1 µJ and repetition rates up to 2 MHz.

*THz Generation:* THz radiation was generated through optical rectification in a 500 µm-thick (E)-4-((4-nitrobenzylidene)amino)-N-phenylaniline (PNPA) organic crystal (THz Innovations), phase-matched for 1250-2100 nm with an optimal phase-matching wavelength of 1550 nm. The polarization of the pump beam was aligned with the c-axis of the crystal, which is orthogonal to the direction of beam propagation for generating THz with the highest possible efficiency. A stepping motor-controlled delay line (VXM-1, Velmex) was used to scan the THz pulse over 3-100 ps with 2.5 µm/step increments. Low pump pulse energy of ~ 1 µJ and an organic crystal (PNPA) were utilized to generate THz electric fields with a THz generation efficiency exceeding 4%.[23]

*Electro-Optic Modulation:* A 2 mm thick ZnTe crystal (ZnTe-2000H, Eksma Optics) with <110> orientation was used for probe modulation using electro-optic effect. This crystal offers optimal phase-matching conditions within the desired THz bandwidth (100 GHz-2 THz). Both the THz and probe beams were polarized along the z-axis (001) of the ZnTe crystal for efficient detection of THz. This modulation technique was used for both conventional delay-line-based and single-shot detection modes.

*Delay Line THz Detection (Balanced Photodetection):* A polymer zero-order quarter-wave plate at 780 nm (05RP34-780, Newport) and a Wollaston prism (Calcite crystal, 10WLP08AR.16, Newport) were used to implement balanced photodetection. The elliptically polarized modulated probe beam was split into two orthogonally polarized components, which were then analyzed by a balanced photodetector (PDB210A, Thorlabs). The balanced photodetector had a bandwidth of 1 MHz within a wavelength range of 320 – 1060 nm with a fixed common mode gain of 500 kV/A. To enhance signal-to-noise ratio further, a lock-in amplifier (Moku-go, Liquid Instruments) was employed with a 12 dB/octave roll-off, 30 ms time constant, and 10 dB gain. The resulting signal was acquired using a multi-channel digital oscilloscope (MSO44, Tektronix), and the mean value was extracted at each delay line step using custom software. The motorized delay line (VXM-1, Velmex) was stepping at 133 fs/step.

*Single-Shot THz Detection (Spectral Modulation):* An electro-optic crystal was used to phase-modulated the frequency components of the chirped probe pulse in proportion to the local transmitted THz electric field. This phase modulation was converted into intensity modulation of the probe pulse using a pair of cross-polarizers (LPNIRB100, Thorlabs) with an extinction ratio[17] of > 1000:1 within the range of 695-1100 nm. The spectrometer then analyzed the intensity-modulated chirped probe beam to extract the THz signal. (Procedure detailed in the section 'Data Acquisition and Post-Processing' and SI Sec. 4).

The crossed-polarizer configuration was integrated with a 0.5 m spectrometer (iHR550, Horiba Scientific) and an EMCCD camera (Synapse) for spectral modulation. An 1800 grooves/mm grating (with a blaze wavelength of 500 nm) was used in the spectrometer, with a slit width of 1000 µm and no EM gain. The wide slit width enhanced the SNR for both unmodulated and modulated probe pulses.

*Supercontinuum Generation:* A 6 mm thick sapphire crystal, AR-coated for 780 nm, was used as the bulk medium for supercontinuum generation. Supercontinuum was produced by focusing an uncompressed, negatively-chirped probe beam (350 fs, SI Sec. 3) onto this crystal. A pair of AR-coated plano-convex lenses (f = 1 cm, f/D = 1.67, Thorlabs LA4280-B) focused the 2.5 mm wide ($1/e^2$) laser beam through the crystal and collimated it afterward. A peak probe power of 4 MW was used (greater than the required threshold of 3 MW for supercontinuum generation).[31]

*Pulse Chirping:* Probe pulses were chirped to the desired pulse widths using highly dispersive SF11 glass (SCHOTT, dispersion coefficient: -603.91 ps/(nm-km) at 780 nm) in both single-shot schemes. A 5 cm long SF11 rod was used for dispersing/chirping the supercontinuum beam to 2 ps, while a 10 cm long rod was employed for chirping the probe pulse with limited bandwidth to 1.9 ps. Different thicknesses were chosen to accommodate the differing bandwidths among the probe beams (Fig. 2, SI Sec. 3).

*Pulse Characterization Methods:* A 0.5 m spectrometer (iHR550, Horiba Scientific) was used to characterize the spectral bandwidths of the probe and the supercontinuum beams. It was equipped with an 1800 grooves/nm grating (blaze wavelength: 500 nm) and an EMCCD camera (Synapse). The spectrometer had a spectral resolution of 0.025 nm and a dispersion of 1.34 nm/mm.

The probe was used in three different modes: an unaltered probe beam from an OPA (350 fs), chirped probe beam with limited bandwidth (22 nm, 1.9 ps), and chirped supercontinuum probe beam (45 nm, 2 ps) (SI Sec. 3). A streak camera (Hamamatsu C10910 with 0.1 ns to 50 ns fast sweep unit, M10912-01) with approximately 0.9 ps temporal resolution and an autocorrelator (Thorlabs femtosecond interferometric autocorrelator) with 1 ps measurement range were employed to measure the temporal pulse widths of these pulses (Fig. 2, SI Sec. 3). The streak camera has an MCP gain of $3 \times 10^3$, 6.5 µm pixel size, and a frame rate of 100 frames/s. Jitter correction over 25 - 50 pulses was used to improve the SNR of the temporal pulse shape.

*4.2 Test Section Design*

*Plasma Reactor:* An inductively coupled plasma was generated inside a 1" diameter quartz tube with a 10 cm discharge length. The plasma generator (cito 2710-ACNA-N37A-FF) was connected to a custom impedance matching network and antenna with a 1" coil diameter and 5 turns. The generator was rated for 1000 W at 27.12 MHz. The reactor was designed with a flow configuration (flow rate: 10 sccm) and z-cut quartz windows (3 mm thick) at the ends to minimize THz losses. Argon (Praxair, ~99.999% purity) was allowed to purge the reactor from one end, while a vacuum pump (Edwards nXDS15iC) was used to evacuate the reactor and maintain a pressure of approximately 220 mTorr, as measured by a capacitive pressure gauge (MKS Baratron, 10 Torr range).

*Purge Box:* A purge box was constructed to enclose the entire THz beam path and control the amount of water vapor. The purge box was equipped with a hygrometer to measure the relative humidity within the enclosure. Dry nitrogen ($N_2$) was used to purge the air inside the box, reducing the relative humidity to < 1%, which enabled a background THz signal measurement ($I_0$). Additionally, the purge box allowed for the introduction of various gases into the pump beam path, facilitating the characterization of their absorption spectra.

*4.3 Data Acquisition and Post-Processing*

*Delay Line THz Extraction:* THz transmission through the test section was measured at different time delays by translating a motorized stage over the stage range where the probe and THz pulses overlapped. A balanced photodetector measured the difference in intensities of the cross-polarized probe beams after the Wollaston prism.[24] This data was sent to a lock-in amplifier to improve the SNR. The average lock-in data collected at different stage positions of the THz signal was used to reconstruct the full THz waveform in the time domain. The motorized delay line could step in the range of 50 - 133 fs/step, depending on the required time resolution, with a travel range of up to 30,000 steps.

An FFT was applied to the THz signal to obtain the amplitude and phase spectra in the frequency domain. The phase spectra wrapped around $\pm\pi$, causing discontinuities due to phase jumps. A standard phase unwrapping algorithm was used to correct these discontinuities.[32]

*Single-Shot THz Extraction:* THz transmission through the test section was measured as intensity modulations on a chirped probe pulse. The time history of THz transmission was captured with a single shot of a laser pulse without scanning the delay line. The intensity spectra of the unmodulated and modulated probe beams were measured separately using the spectrometer over a broad range of 720 - 820 nm, depending on the bandwidth of the probe pulse. The desired THz signal was extracted by subtracting the unmodulated probe intensity spectrum from the modulated probe spectrum and normalizing the result with the unmodulated probe spectrum (SI Sec. 4). Since the spectrometer output was in wavelength, it was converted to time using a time-wavelength calibration procedure (76 fs/nm for chirped pulse and 42 fs/nm for supercontinuum) (SI Sec. 4). Finally, an FFT was applied to the extracted THz signal to obtain the amplitude and phase spectra.

*4.4 Models*

*Spectral Encoding:* A spectral encoding and decoding model was used to validate, generalize, and extrapolate the experimental results. This model simulates how the spectral features of the chirped probe pulse are modulated by the THz electric field, through electro-optic modulation. It also predicts how the spectrometer detects and interprets this modulated signal. The transmitted modulated probe pulse, $E_m(t)$, that makes it through the ZnTe crystal (i.e., the electro-optic modulation crystal) is,

$$E_m(t) = E_u(t)[1 + kE_{THz}(t)], \qquad (1)$$

where $E_u(t)$ is the unmodulated chirped probe pulse, and $E_{THz}(t)$ is the THz electric field (i.e., the signal of interest).[20] k ($|k| \ll 1$) is the modulation constant. It is dependent on many factors within an optical setup including the amount of light scattering in a detection crystal, the thickness of the crystal, its electro-optic coefficient, group velocity mismatch, and optical bias.[17] As a spectrometer measures intensity, $I_m$ and $I_u$, in spectral domain, it can be shown that (SI Sec. 1),

$$E_{THz}(t) \propto \frac{I_m(\omega) - I_u(\omega)}{I_u(\omega)}. \qquad (2)$$

The spectral function of the spectrometer was assumed to be a delta function because a spectrometer resolution ($\delta\omega_s$) is generally much lower than the single-shot measurement frequency resolution (i.e., $\delta\omega_s \ll \frac{T_{THz}}{T_c T_0}$).

The THz electric field waveform, $E_{THz}(t)$, in Eq. 1 was simulated using two approaches based on the context: (1) For comparing experimental and simulated results, the THz waveform that was measured experimentally using the delay line method was used. (2) For generalization or extrapolation beyond experimental conditions, a bipolar signal (Eq. 3) was used,

$$E_{THz}(t) = \frac{t}{T_{THz}} \exp\left(-\frac{t^2}{T_{THz}^2}\right), \qquad (3)$$

where $T_{THz}$ is THz pulse width. $E_u(t)$ is a Gaussian probe pulse chirped to a desired pulse width (or measurement window). $E_m(t)$ in time domain was extracted using these two signals (i.e., $E_{THz}$ and $E_u$) using Eq. 1. To compute the spectral domain modulated and unmodulated probe electric fields (i.e., $\tilde{E}_m(\omega)$ and $\tilde{E}_u(\omega)$ respectively), an FFT algorithm was employed. The modulation constant k was assumed to be 0.01 because $|k| \ll 1$, and its exact value is not critical as the final THz electric field amplitude is normalized. Based on the spectrometer's resolution of 0.025 nm and a spectral measurement range of 100 nm, a total measurement time window of 200 ps (spanning from -100 ps to 100 ps) and a sampling period of 0.5 fs were used in the simulations. Since spectrometers measure intensity rather than electric fields, the electric fields $\tilde{E}_m(\omega)$ and $\tilde{E}_u(\omega)$ were squared to convert them into intensity values ($I_m(\omega)$ and $I_u(\omega)$)) (SI Sec. 7).

$$I_{m(u)}(\omega) = \left\|\tilde{E}_{m(u)}(\omega)\tilde{E}^*_{m(u)}(\omega)\right\| \text{ where } \tilde{E}_{m(u)}(\omega) = \int_{-\infty}^{\infty} E_{m(u)}(t) \exp(-i\omega t)\, dt, \qquad (4)$$

which is valid for modulated m or unmodulated (u) probe pulses.

All the forms of probe pulses (e.g., unchirped, chirped, and supercontinuum) were modeled as Gaussian pulses, each characterized by distinct transform-limited pulse widths ($T_0$). Notably, the $T_0$ for the supercontinuum is shorter than that of the standard probe pulse, simulating the broader bandwidth of the supercontinuum compared to the standard probe pulse. Apart from this difference in $T_0$, both supercontinuum and chirped pulse were simulated to using the same spectral encoding algorithm.

*Beer-Lambert Law Fit*: Quantitative absorption spectroscopy was calculated using single-shot THz-TDS by fitting Beer's law to the experimentally measured absorbance,

$$A(\omega) = \log\left(\frac{I_0(\omega)}{I(\omega)}\right) = \sigma(\omega)L\tilde{n}, \quad (5)$$

where ω refers to the spectral frequency, $I_0(\omega)$ refers to the original intensity at a given frequency, $I(\omega)$ refers to the transmitted intensity at that frequency, σ refers to the absorption coefficient, L is the length of the gas column and ñ is line-averaged column number density of the species under consideration. $I_0$ and I were measured experimentally, while the HITRAN database[33] provided σ, Lorentzian air-broadened and self-broadened FWHMs, and pressure shifts. These values were used to fit σ with a Voigt profile. The fit was used to analyze THz absorbance data for a given L to extract ñ. The gas temperature was maintained at room temperature, and the pressure remained constant at 1 atm throughout the measurement.

*Absorbance Models*: THz absorbance features were extrapolated to longer $T_c$ and $T_0$ values using the spectral encoding algorithm. The process began with a bipolar signal (Eq. 3) with a $T_{THz}$ of 0.25 ps, representing the THz pulse generated through optical rectification, as the original input (SI Sec. 7). To calculate absorbance (Eq. 5), both the transmitted intensity (I) and the original unmodulated intensity ($I_0$) are required (Eq. 5). To extract $I_0$, the spectral encoding algorithm was applied to the bipolar signal to extract the single-shot THz signal (as seen by a chirped or supercontinuum pulse). An FFT was then performed on the resultant THz electric field, and the square of the amplitude of this resultant electric field in the frequency domain provided the THz intensity spectral profile ($I_0$).

The transmitted THz intensity (I) was calculated with one additional step. The HITRAN database was employed to simulate the THz signal after propagating through an absorbing medium (in this case, water vapor). Absorption was calculated using Beer's law in the frequency domain. The detailed step-by-step process for this algorithm is provided in SI Sec. 7.

*Drude Model*: Two different schemes were used to extract plasma properties using SETS with the help of Drude model[19,27] in two different forms: 1) Under specific conditions where $\omega \gg \nu_{en}, \omega_p$ (SI Sec. 5), a simplified Drude model (Eq. 6-7) was used to fit the experimental phase difference/phase shift (Δϕ) between the THz pulses, with and without plasma, to extract $\tilde{n}_e$. Δϕ was determined by performing a Fourier transform on each pulse individually and then calculating the difference between the phases of these Fourier transforms.

$$\Delta\phi(\omega) = \frac{\omega}{c}\int_0^L (1-N(\omega))\,dx = \frac{\tilde{n}_e e^2}{2c\varepsilon_0 m_e \omega}, \quad (6)$$

$$\tilde{n}_e = \frac{1}{L}\int_0^L n_e dx, \quad (7)$$

where c is speed of light, L is plasma length, $m_e$ is the mass of an electron, $\varepsilon_0$ is permittivity of free space, e is the charge of an electron, and N is the refractive index of the plasma medium. 2) The complete Drude model (Eq. 8-10) was applied to fit the experimental time domain plasma-dispersed THz signal, enabling the simultaneous extraction of both $\omega_p$ and $\nu_{en}$ in a longitudinally-uniform plasma (SI Sec. 5-6).

$$E_{THz,\,plasma}(t) = \text{Re}\left(\frac{1}{2\pi}\int_{-\infty}^{\infty} \tilde{E}_{THz,\,plasma}(\omega)\exp(i\omega t)\,d\omega\right), \quad (8)$$

$$\widetilde{E}_{THz, plasma}(\omega) = \widetilde{E}_{THz}(\omega) \exp\left(\frac{i\omega L N(\omega)}{c}\right), \quad (9)$$

$$N(\omega) = \sqrt{1 - \frac{\omega_p^2}{\omega(\omega + i\upsilon_{en})}} . \quad (10)$$

where $E_{THz, plasma}(t)$ is the simulated THz electric field after passing through plasma, $\widetilde{E}_{THz}(\omega)$ is the undispersed THz signal inverted into the frequency domain using FFT, $\widetilde{E}_{THz, plasma}(\omega)$ is the THz-signal transformed by the plasma in the frequency domain (SI Sec. 5, Fig. S6).

*4.5 Statistical analysis*

The experimental results presented in Figs. 3 and 4 were obtained by averaging separate measurements for each corresponding signal/data point (N = 5).

## Acknowledgments

**Funding:** This research is supported by the Office of Naval Research grant N00014-23-1-2306, with Ryan Hoffman as Program Manager.

**Author contributions:** CRN and TU conceived the research; CRN designed the experiments and acquired the data; CRN and TU analyzed the data; CRN wrote the manuscript; CRN and TU revised the manuscript.

**Competing interests:** Authors declare that they have no competing interests.

**Data and materials availability:** All data are available in the main text or the supplementary materials.

**Supplementary Materials**

Supplementary Text

Figs. S1-S10